\documentclass[english,aps,prd,superscriptaddress,preprintnumbers,nofootinbib,preprint]{revtex4}
\usepackage[T1]{fontenc}
\usepackage[latin9]{inputenc}
\usepackage{verbatim}
\usepackage{amsmath}
\usepackage{dcolumn}
\usepackage{graphicx}
\newcommand{\be}{\begin{equation}}
\newcommand{\ee}{\end{equation}}
\newcommand{\bea}{\begin{eqnarray}}
\newcommand{\eea}{\end{eqnarray}}


\usepackage{babel}
\usepackage{color}
\begin{document}
\global\long\def\order#1{\mathcal{O}\left(#1\right)}
\global\long\def\d{\mathrm{d}}
\global\long\def\P{P}
\global\long\def\amp{{\mathcal M}}
\preprint{TTP15-035, CERN-PH-TH-2015-226, FERMILAB-PUB-15-398-T}

\def\FNAL{Department of Theoretical Physics, Fermilab, Batavia, IL, USA}
\def\KIT{Institute for Theoretical Particle Physics, KIT, Karlsruhe, Germany}
\def\CERN{CERN Theory Division, CH-1211, Geneva 23, Switzerland}

\title{ 
QCD corrections to $ZZ$ production in gluon fusion at the LHC}

\author{Fabrizio Caola}            
\email[Electronic address: ]{fabrizio.caola@cern.ch}
\affiliation{\CERN}

\author{Kirill Melnikov}            
\email[Electronic address: ]{kirill.melnikov@kit.edu}
\affiliation{\KIT}

\author{Raoul R\"ontsch }            
\email[Electronic address: ]{rontsch@fnal.gov}
\affiliation{\FNAL}

\author{Lorenzo Tancredi }            
\email[Electronic address: ]{lorenzo.tancredi@kit.edu}
\affiliation{\KIT}

\begin{abstract}
We compute the next-to-leading order  QCD corrections to the  production of 
two $Z$-bosons in the annihilation of two gluons at the LHC.  Being enhanced by a large 
gluon flux, these corrections 
provide distinct and, potentially, the dominant   part of the N$^3$LO QCD contributions  
to $Z$-pair production in proton collisions. 
The  $gg \to ZZ$ annihilation  is a loop-induced process 
that receives the dominant contribution from loops of five light  quarks, that are  included 
in our computation in the  massless approximation. 
We find that  QCD 
corrections increase the $gg \to ZZ$ production cross section by  ${\cal O}(50\%-100\%)$
depending on the values of the renormalization and factorization scales used in the leading 
order computation, and the collider energy.  The large corrections to $gg \to ZZ$ channel increase the  
$pp \to ZZ$ cross section  by about six to eight percent, exceeding   the estimated theoretical 
uncertainty of the recent NNLO QCD calculation. 
\end{abstract}

\maketitle

\section{Introduction} 
Production of pairs of vector bosons in proton collisions is  one of the most interesting processes  
 studied by ATLAS and CMS  during the  LHC Run I \cite{atlas7,cms7,cms8}.
  Indeed, $pp \to ZZ$, $pp \to  W^+W^-$, 
and $pp \to \gamma \gamma$ were  instrumental for the discovery of the Higgs boson.
As the focus of  Higgs physics shifts  from the discovery to precision 
studies of the Higgs boson properties,  di-boson production 
 processes become essential for  constraining   anomalous Higgs boson couplings, for measuring the 
quantum numbers of the Higgs boson and for studying the Higgs boson width, 
see Refs.~\cite{Khachatryan:2014jba,Khachatryan:2014kca,Khachatryan:2014iha,atlaswidth}.
Additionally, these processes provide important tests 
of our understanding of the Standard Model and can be used to constrain  anomalous 
electroweak  gauge boson couplings.

Production of electroweak gauge boson pairs occurs mainly due to quark-antiquark annihilation 
$q \bar q \to V_1 V_2$. 
This contribution is known through next-to-next-to-leading order (NNLO) in perturbative QCD
\cite{Catani:2011qz,Grazzini:2013bna,Cascioli:2014yka,Gehrmann:2014fva,Grazzini:2015nwa,Grazzini:2015hta}.
However, as was pointed out in 
Refs.~\cite{Glover:1988rg,Glover:1988fe,Dicus:1987dj}, 
there is a sizable contribution from the gluon annihilation channel 
$gg \to V_1 V_2$,  whose significance depends on the selection  cuts.  
For example, 
aggressive cuts applied to $pp \to W^+W^-$ to separate the Higgs boson signal from the continuum 
background can increase 
the fraction of gluon fusion events in the background sample \cite{Binoth:2006mf}.
Since $gg \to V_1 V_2$ is a one-loop process and since production of 
electroweak boson pairs at leading order (LO) occurs only in the $q \bar q$ channel, the gluon fusion contribution 
to $pp \to V_1 V_2$ through NNLO  only needs to be known at  leading order, i.e. the one-loop   approximation. 
Thus, all existing numerical  estimates of the significance of the
gluon fusion mechanism in weak boson pair production ignore radiative corrections 
to $gg \to ZZ$ that are, potentially,  quite large~\cite{Bonvini:2013jha}.  The need 
to have an accurate estimate of QCD corrections to gluon fusion processes for  the Higgs 
width~\cite{Caola:2013yja,Campbell:2013una} and generic off-shell 
measurements \cite{Kauer:2012hd,ellis,Azatov:2014jga} 
was  strongly emphasized   in Ref.~\cite{atlaswidth}.

In this paper, we will focus on the calculation of the next-to-leading order (NLO) QCD corrections to the gluon fusion 
contribution to $pp \to ZZ$ process. 
The largest contribution to $gg \to ZZ $ comes from quarks of the first two generations;
these quarks can be taken to be massless.  The situation is  more complicated for quarks 
of the third generation. Ideally, we would like to include the (massless) bottom quark  contribution 
and ignore the contribution of the massive top quark  since,  
at leading order, the top-quark contributions change the cross section 
by only about $1\%$ (cf. Refs.~\cite{mcfm,gg2VV}).\footnote{ Contribution
of the top quark loop  becomes non-negligible in the region of high  four-lepton invariant 
masses $m_{4l} > 2 m_t$.} 
We can separate bottom and top contributions everywhere except in  triangle diagrams that involve  
anomalous correlators of vector and axial currents.  
In these  triangle diagrams, when bottom and top contributions are 
combined, the residual 
contributions are  suppressed by the top quark mass, provided that we 
can assume it  to be 
larger than any other energy scale in the problem. Unfortunately,  in these  diagrams top and bottom 
contributions can not be separated because the resulting theory is anomalous. 
To deal with this issue, we adopt the  following strategy: we include  
quarks of the first two generations and the $b$-quark in our calculation in 
the massless approximation and we neglect 
all triangle diagrams   whose contribution is then naturally associated 
with the quark contributions to $gg \to ZZ$ process.
We note that  the evaluation of the NLO QCD
corrections to top quark mediated contribution to $gg \to ZZ$ process 
is not yet possible because the relevant two-loop amplitudes are not available. However,  
such contributions  were recently studied in 
Ref.~\cite{Melnikov:2015laa} in the approximation of a very large mass of 
the top quark. In that calculation quite large QCD corrections were found.

Computing NLO QCD corrections to $gg \to ZZ$ process  is challenging because it is loop-induced. 
 For this reason, the NLO QCD computation 
requires two-loop virtual matrix elements  for $gg \to ZZ$ and 
 one-loop matrix elements for $gg \to ZZg$ processes.  
The recent progress in calculating  two-loop integrals with two massless and two massive external lines 
\cite{Gehrmann:2013cxs,Gehrmann:2014bfa,planar,nonplanar,Papadopoulos:2014hla} 
made it possible to compute the required two-loop scattering amplitudes.  
Such amplitudes  were  calculated  recently for  
$q\bar q \to V_1 V_2 $~\cite{Caola:2014iua,Gehrmann:2015ora}
and 
$gg \to V_1 V_2 $~\cite{Caola:2015ila,vonManteuffel:2015msa}  
processes.

The second ingredient that we need is the $gg \to ZZ g$ amplitude. Since this is a one-loop 
amplitude, it can be calculated in a  relatively standard way, at least as a matter of principle. In fact, such 
calculations were performed  in the past~\cite{Agrawal:2012df,Campanario:2012bh}
and used to predict the production cross section 
for  $pp \to ZZ+j$. Automatic tools for one-loop computations 
 can also deal with this process~\cite{openloops,madloop}.
Nevertheless, it is a non-trivial computation since, if we aim at calculating 
the NLO QCD corrections to $gg \to ZZ \to 4l$,   we require 
fast and stable  
calculation of helicity amplitudes for $ gg \to ZZ g$ process 
that  includes  decays of $Z$-bosons to leptons 
and  can be extrapolated 
to soft and collinear kinematics of the final state gluon. Because of that, we decided 
to construct our own implementation of the scattering amplitude for $gg \to ZZg$ using the unitarity methods 
\cite{bern,britto,opp,Ellis:2007br,me}.\footnote{For recent  reviews see  
Refs.~\cite{Ellis:2011cr,Henn:2014yza}.}

The paper is organized as follows. In Section~\ref{section1} we present a 
brief review of the calculation 
of  the two-loop scattering amplitude for $gg \to ZZ$ process.
In Section~\ref{section2} we discuss the calculation of the one-loop 
helicity amplitudes for $gg \to ZZg$ and present numerical results for a kinematic point. 
In Section~\ref{section3} we present  numerical results for $gg \to ZZ$ contribution 
to $pp \to ZZ$ process at $8$ and $13$~TeV LHC at 
leading and next-to-leading order in perturbative QCD.
We conclude  in Section~\ref{section4}. 

\section{The two-loop scattering amplitudes for $gg \to ZZ$}
\label{section1}
We start with a brief  discussion of the two-loop scattering amplitudes for $gg \to ZZ$ process. 
Helicity amplitudes for this process were recently computed  in Refs.~\cite{Caola:2015ila,vonManteuffel:2015msa}. 
In these references, each of the two independent helicity amplitudes for the process $gg \to ZZ \to 4l$ was 
written as linear combinations of nine form factors that depend on the Mandelstam invariants 
of the ``prompt'' process $gg \to ZZ$ and the invariant masses of the two $Z$ bosons.  The form factors 
are expressed in terms of polylogarithmic functions, including both ordinary and Goncharov polylogarithms. 

In this paper we use the results of Ref.~\cite{vonManteuffel:2015msa} which are implemented in a C$++$ code
that can produce numerical results with arbitrary precision. In order to detect possible numerical 
instabilities, the code compares numerical evaluations obtained with different 
(double, quadruple and, if required, arbitrary) precision settings. 
If the results differ beyond a chosen 
tolerance, the precision is automatically increased. 
Of course,  switching to arbitrary precision increases the evaluation time  substantially.
Fortunately, we found that  for phenomenologically relevant  situations, 
  the  number  of points  where the code 
switches  to arbitrary precision  is negligible.
Such  points originate from kinematic regions where the two $Z$-bosons have either vanishing 
kinetic energies or vanishing transverse momenta. The amplitude squared is integrable in both of these regions, 
but, in practice, it can become numerically  unstable.  Since the contribution of these 
regions to the $gg \to ZZ$ cross section is relatively small, 
cutting them away, in principle, leads to an opportunity to perform 
stable numerical integration of the two-loop virtual correction over the four-lepton phase-space,
resorting to quadruple precision only. However, we found that
the improvement in performance achieved by  cutting away the problematic 
 regions is rather  limited, so we used
the default arbitrary precision implementation of the two-loop amplitude in practice.

Since the $gg \to ZZ$ amplitude is one of the most complicated amplitudes that are currently known 
analytically, 
it is interesting to point out that the required evaluation times 
are  acceptable for phenomenological needs. Indeed, calculation of all helicity 
amplitudes requires  about two seconds 
per phase-space point in quadruple precision and, since 
 the phase-space for $gg \to ZZ$ is relatively simple, 
one does not need excessively large  number 
of points  to sample it with good precision.

For further reference we provide numerical results for the finite
remainder of the one- and two-loop scattering amplitudes
defined in $q_t$-subtraction scheme, see 
Ref.~\cite{vonManteuffel:2015msa}.
The numerical results are presented for the choice of the renormalization  scale $\mu=\sqrt{s}$, where 
s is  the partonic  center-of-mass energy squared.  
The $q_t$-subtraction scheme~\cite{cagr}  is discussed in detail   in Ref.~\cite{Catani:2013tia}.
We consider the kinematical point
$$g(p_1) + g(p_2) \to (Z/\gamma)(p_{34}) + (Z/\gamma)(p_{56}) \to e^-(p_3) + e^+(p_4) + \mu^-(p_5) + \mu^+(p_6) $$
with (in GeV units)
\begin{small}
\be
\begin{split} 
     &  p_1 =   ( 99.5173068698129, 99.5173068698129,0,0), \\
     &  p_2 = ( 99.5173068698129, -99.5173068698129,  0, 0 ), \\ 
     &  p_3 = ( 45.1400347869485, 43.4878610174890 ,-9.85307698310431, 7.02463939683013 ), \\ 
     &  p_4 = ( 55.6586029753540, -27.4053916434553,48.1951275617684 , 4.90451560725290), \\
     &  p_5 = ( 36.2015682945089, 34.5902512456859, -8.01242197258994, 7.06180995747356), \\ 
     &  p_6 = (  62.0344076828144,-50.6727206197196,-30.3296286060742, -18.9909649615566 ), \\
\end{split} 
\ee
\end{small}
and define a normalized amplitude through the following equation 
\be
d\,\sigma_{gg \to (Z/\gamma)(Z/\gamma) \to 4l} = \frac{(N_c^2-1)}{512 s}
\times 10^{-6} \times \sum_{\lambda_1,\lambda_2,\lambda_e,\lambda_\mu} 
\left| A(1_g^{\lambda_1},2_g^{\lambda_2}; 
3_{e^-}^{\lambda_e},4_{e^+}^{-\lambda_e},5_{\mu^-}^{\lambda_\mu}, 6_{\mu^+}^{-\lambda_\mu}) \right|^2  
d\,{\rm LIPS_{4} }. 
\label{eqdef}
\ee
Note that  in Eq.(\ref{eqdef}) all the 
color factors have been factored out and $d {\rm LIPS}_4$ is the standard Lorentz-invariant 
phase-space of the four final leptons. The color-stripped amplitude admits  an expansion 
in the strong coupling constant 
\begin{align}
A(1_g^{\lambda_1},2_g^{\lambda_2};& 
3_{e^-}^{\lambda_e},4_{e^+}^{-\lambda_e},5_{\mu^-}^{\lambda_\mu}, 6_{\mu^+}^{-\lambda_\mu}) =
\left(\frac{\alpha_s(\mu)}{2 \pi}\right) 
\left[ {\cal A}_{1l}(1_g^{\lambda_1},2_g^{\lambda_2}; 
3_{e^-}^{\lambda_e},4_{e^+}^{-\lambda_e},5_{\mu^-}^{\lambda_\mu}, 6_{\mu^+}^{-\lambda_\mu}) \right.
\nonumber \\
&\left. +\left(\frac{\alpha_s(\mu)}{2 \pi}\right) 
 {\cal A}_{2l}(1_g^{\lambda_1},2_g^{\lambda_2}; 
3_{e^-}^{\lambda_e},4_{e^+}^{-\lambda_e},5_{\mu^-}^{\lambda_\mu}, 6_{\mu^+}^{-\lambda_\mu}) + 
\mathcal{O}\left(\alpha_s^2\right)
 \right]\,.
\end{align}
Numerical results for the two independent helicity amplitudes at one- and two-loops 
 are given in Table~\ref{table1}. We emphasize that the results in Table~\ref{table1} are given in 
the $q_t$-subtraction scheme, c.f. Ref.~\cite{vonManteuffel:2015msa}.

\begin{table}[t]
\begin{center}
\begin{tabular}{|l|c|c|}
\hline
Helicity amplitude &  $1$-loop  & $2$-loop    \\
\hline
$\;{\cal A}({1^-},{2^-};{3^-},{4}^+,{5^-},{6}^+)$ &  
$-3.6020208 - 0.80680028~i$
& 
$-87.785548 +35.086257~i$
   \\
\hline
$\;{\cal A}({1^-},{2^+};{3^-},{4}^+,{5^-},{6}^+)$ & 
$+0.2507409 + 0.38426042~i$
 & 
$+18.585086  + 7.5961902~i$
   \\
\hline
\end{tabular}
\end{center}
\caption{Results (in GeV$^{-2}$) for normalized $q_t$ remainder of one- and two-loop amplitudes for different 
choices of gluon and lepton helicities, evaluated at the scale $\mu=\sqrt{s}$. See text for details.} 
\label{table1}
\end{table}

\section{The one-loop scattering amplitude  $0 \to gggZ Z$} 
\label{section2}

\begin{figure}
\centering
\includegraphics[width=0.8\textwidth]{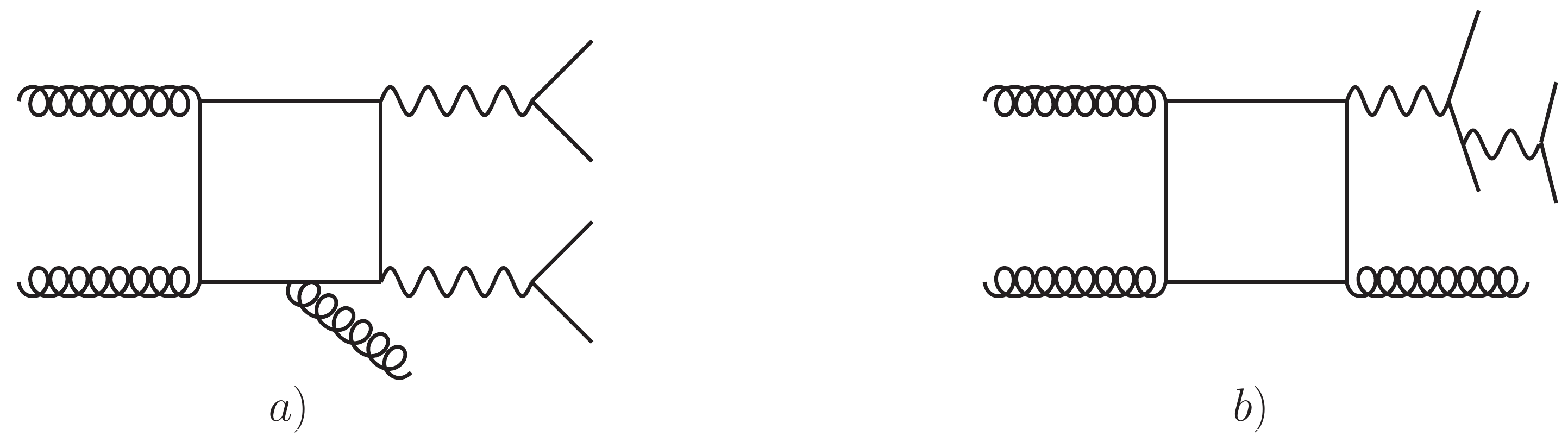}
\caption{Representative Feynman diagrams for the $0\to gggZ(\to e^-e^+)Z(\to\mu^-\mu^+)$ amplitude.
 \emph{Double resonant} diagrams (a) are relevant
for both the on-shell and the 
off-shell production. \emph{Single resonant} diagrams (b) are only relevant for the off-shell production and
are not included in our computation. See text for details.}\label{feyndia}
\end{figure}

In this Section, we discuss the computation of the one-loop scattering amplitude required
for the calculation of the inelastic process $gg \to ZZ+g$.\footnote{To simplify the notation, in this section we do not consider 
photon-mediated four-lepton amplitudes. For phenomenological results discussed in Sec.~\ref{section3}, we consider the full $gg\to (Z/\gamma)(Z/\gamma)+g$ amplitude.}
To this end, we  consider 
the process $0 \to g(p_1) g(p_2) g(p_3) Z(p_{45}) Z(p_{67})$. 
Decays of the $Z$-bosons are allowed but, since we are interested in the on-shell production 
of the two $Z$-bosons,  we do not include single resonant diagrams 
where one of the $Z$-bosons is emitted from the decay products of the other one, see Fig.~\ref{feyndia}.
We will refer to the decay  products of the $Z$-boson with momentum $p_{45}$ as the electron and the positron 
with momenta $p_{4}$ and $p_5$ and to the decay products of the $Z$ boson with momentum $p_{67}$ 
as the muon and the anti-muon with momenta $p_{6}$ and $p_7$, respectively.  
All leptons are taken to be massless. 
Since helicities of massless leptons are conserved, we only need to specify helicities of the final 
state leptons $e^-$ and $\mu^-$; the allowed helicities of the positron and the anti-muon in the final 
state are then automatically fixed. 

We write the interaction vertex of the $Z$-boson and a fermion pair as 
\be
Z \bar f \gamma_\mu f  \in g_{L,f} \frac{\gamma_\mu( 1 + \gamma_5)}{2} 
 + g_{R,f} \frac{\gamma_\mu ( 1 -  \gamma_5)}{2}, 
\;\;\;\;  f \in (l,q).
\ee
The left and right couplings for leptons and quarks are given by an identical  formula
\be
g_{L(R),f} = \frac{V_f \pm A_f}{\cos \theta_W},
\ee
where we use {\it i}) $V_l = -1/2 + 2 \sin^2 \theta_W$, $A_l = -1/2$ for charged leptons;
{\it ii}) $V_u = 1/2 - 4/3 \sin^2 \theta_W$, $A_u = 1/2$ for up-type  quarks; 
and {\it iii}) $V_d = -1/2 + 2/3 \sin^2 \theta_W$,
$A_d = 1/2$ for down-type  quarks. 

The $0 \to ggg ZZ$ scattering  amplitude can be written  as a sum of two terms
\be
{\cal A}^{ZZ} = g_s^3 g_W^4 \left({\rm Tr} \left [  t^{a_1} t^{a_2} t^{a_3}  \right ] A_{123}^{ZZ} +  
{\rm Tr} \left [  t^{a_1} t^{a_3} t^{a_2}  \right ]  A_{132}^{ZZ}\right), 
\ee
with ${\rm Tr}(t^a\;t^b)=\delta^{ab}/2$. 
The two color-ordered amplitudes, stripped of their couplings to leptons and quarks, 
are defined as  
\be
\begin{split}
& A_{ijk}^{ZZ} = 
 C_{\lambda_e,e} C_{\lambda_\mu, \mu} \left ( g_{LL}^{ZZ} 
{\cal A}^{LL}_{ijk}(\lambda_i, \lambda_j, \lambda_k; \lambda_e, \lambda_\mu)  
+
g_{RR}^{ZZ} 
{\cal A}^{RR}_{ijk}(\lambda_i, \lambda_j, \lambda_k; \lambda_e, \lambda_\mu) 
\right ). 
\end{split}
\label{eq45} 
\ee
In Eq.(\ref{eq45}) we introduced 
\be
C_{\lambda,l} = D_Z(m_{ll}^2) \left ( g_{L,l} \delta_{\lambda,-}  + g_{R,l} \delta_{\lambda,+} \right ),
\label{eq1}
\ee
where $D_Z(s)$ is the function related to 
the Breit-Wigner propagator $D_Z(s) = s/(s - M_Z^2 + i M_Z \Gamma_Z)$. 
The couplings $g_{LL}^{ZZ}$ and $g_{RR}^{ZZ}$  are expressed through  $Z$-boson couplings to quarks 
propagating in the loops 
\be
g_{LL}^{ZZ} =  \sum \limits_{q} g_{L,q}^2 ,
\;\;\;\;
g_{RR}^{ZZ} =   \sum \limits_{q} g_{R,q}^2.
\ee

Given  these definitions,  it is easy to see 
that the helicity amplitudes ${\cal A}^{LL,RR}_{ijk}(\lambda_i, \lambda_j, \lambda_k; \lambda_e, \lambda_\mu)$
can be  calculated for {\it vector} couplings of $Z$-bosons to leptons and quarks,  provided that one 
keeps  left-handed (right-handed) quarks propagating clockwise in the fermion loop. 
This is a natural separation if the scattering amplitudes 
are computed using the unitarity methods \cite{bern,britto,opp,Ellis:2007br,me}.  
There is a useful relation between  left- and right-handed helicity amplitudes  for  two orderings 
of  external gluons
\be
\begin{split} 
& {\cal A}_{132}^{LL}({\lambda_1},{\lambda_3},{\lambda_2}; 
{\lambda_e},{\lambda_\mu})
= - 
{\cal A}_{123}^{RR}({\lambda_1},{\lambda_2},{\lambda_3}; 
{\lambda_e},{\lambda_\mu}),
\\
& {\cal A}_{132}^{RR}({\lambda_1},{\lambda_3},{\lambda_2}; 
{\lambda_e},{\lambda_\mu})
= - 
{\cal A}_{123}^{LL}({\lambda_1},{\lambda_2},{\lambda_3}; 
{\lambda_e},{\lambda_\mu}).
\end{split} 
\ee
These equations suggest that  it is sufficient to compute $LL$ and $RR$ amplitudes for a single ordering; once 
this is done,  all relevant  amplitudes for the second ordering can be constructed.
Finally, we emphasize that we exclude the Breit-Wigner factor\footnote{As we mentioned  earlier, 
we are interested in the on-shell production of the two $Z$-bosons in this paper. However, 
we construct the relevant piece of $gg \to ZZ$ amplitude in full generality, including 
Breit-Wigner propagators for the $Z$-bosons, to enable its later use to study QCD corrections 
to $gg \to ZZ^*$ process.}
 for the 
$Z$-bosons from the definition of the color-ordered helicity amplitudes but we include the $1/s$ factor in its place; this  
can be clearly seen from the definition of the  $D_Z(s)$ function in Eq.(\ref{eq1}).

It is well-known that any one-loop 
amplitude can be  written as a linear combination of one-loop scalar  
integrals that include four-, three- and two-point functions and a rational part 
\be
\label{eq5}
{\cal A}^{LL,RR}_{ijk}({\lambda_i},{\lambda_j},{\lambda_k}; 
{\lambda_e},{\lambda_\mu})
= \sum c_{i}^{LL,RR }  I_i + R^{LL,RR},
\ee
The coefficients $c_{i}$ in the above equation, as well as the rational part, 
can be calculated using unitarity methods.

The idea of the unitarity method is that one can calculate the different discontinuities of the 
left- and right-hand sides of Eq.(\ref{eq5})  and then combine them in such a way 
that coefficients $c_i$ are extracted algebraically.  Calculation of the reduction coefficients 
and the rational part can be performed either analytically or numerically. In this paper, we use a 
mixed approach. We compute the coefficients $c_i$ using numerical four-dimensional unitarity introduced in 
Ref.~\cite{Ellis:2007br}.  The rational part, on the other hand, 
is computed analytically using the method described in Refs.~\cite{direct,
Campbell:2014gua}.  Technical details about the 
unitarity methods used for one-loop computations in  QCD can be found in   Ref.~\cite{Ellis:2011cr}.

\begin{table}[t]
\begin{center}
\begin{tabular}{|l|c|c|}
\hline
Helicity &  LL  & RR    \\
\hline
$\;\tilde{\cal A}_{123}(+,+,+;-,-)$ &  
$-42.714233 + 117.60020~i$
& 
$-138.32358 + 139.68765~i$
   \\
\hline
$\;\tilde{\cal A}_{123}(+,+,-;-,-)$ & 
$+134.26016  + 161.13392~i$
 & 
$+138.09750  + 188.27580~i$
   \\
\hline
$\; \tilde{\cal A}_{123}(-,-,+;-,+)$ &  
$ -32.287418 + 2.1139258~i$
& 
$-31.55258 + 32.433444~i$ \\
\hline
\end{tabular}
\end{center}
\caption{ Results (in GeV$^{-3}$) for normalized  color-ordered amplitudes 
for $0 \to gggZ(e^+e^-)Z(\mu^+\mu^-)$ process, 
for different 
choices of gluon and lepton helicities. See text for details.}
\label{table2}
\end{table}

From the point of view of the  unitarity methods,
the peculiarity of $0 \to gggZZ$ process is that it involves two colorless particles, making 
full color ordering for scattering amplitudes impossible.  This has the following 
implications. Any unitarity computation 
starts with the list of  independent ``parent diagrams'' that are subsequently cut  into 
on-shell scattering amplitudes. Although parent diagrams are independent by construction, 
not all their cuts are, if permutations of external particles are allowed. The challenge,  
therefore,  is to start with the ``parent diagrams'', write down all the cuts that they might have 
and then exclude all the cuts that are not independent.  This issue was successfully 
dealt with  in the context of many recent calculations of one-loop scattering amplitudes 
for quarks, gluons and  
vector  bosons, see e.g.   Refs.~\cite{Ellis:2008qc,
Melia:2010bm,Melia:2011dw,Melia:2012zg}. 
In this paper, we construct the independent set of unitarity cuts following the methodology 
explained in those references.

After identifying independent cuts, we find 39 quadruple, 45 triple and 18 double cuts. There are no 
single-line cuts since internal fermions in our calculation are massless.  
Each of these cuts is described  by a product of tree-level color-ordered amplitudes. The required helicity amplitudes 
include $ \bar q g q$, $\bar q Z q$, $\bar q gg q$, $\bar q gZ q$, $\bar q ZZ q$, $\bar q ggg q$, 
$\bar q gg Z q$, $\bar q g ZZ q$ and $\bar q ggg Z q$. Here, we use a 
generic notion of a $Z$-boson  for an external vector particle  but what we really mean are amplitudes 
with the vector 
current sandwiched between lepton and anti-lepton spinors.  The relevant tree-level amplitudes 
can be extracted from different publications; we have mostly  benefited from a comprehensive 
description  of helicity amplitudes that involve quarks, gluons and vector bosons in Ref.~\cite{badger}. 

The calculation of the rational part of the $0 \to gggZZ$ amplitude is performed analytically, 
using techniques suggested in Ref.~\cite{direct,Campbell:2014gua}. Similar to the cut-constructable 
part the rational amplitude receives contributions from quadruple, triple and 
double cuts. However, for the case of $gg \to ZZg$ amplitudes, the double 
cut contribution vanishes; the rational part  therefore can be 
reconstructed from the calculation of boxes and triangles.  Unfortunately, even 
in this case, the analytic results for the rational part are unwieldy and we choose 
not to present them here. 

For further reference, we give numerical  results for the scattering amplitudes  below.
We consider a kinematic point (momenta are given in GeV)
\begin{small}
\be
\begin{split} 
     &  p_1 =   ( -238.714576090637,-238.714576090637,0,0), \\
     &  p_2 = ( -1021.22119318758, 1021.22119318758 , 0, 0 ), \\ 
     &  p_3 = ( 250.736037681104, -207.896850811885, -124.613643938661, 64.1786550096635 ), \\ 
     &  p_4 = ( 553.889863453468, -495.644737899924, -245.099246845329, 32.5059044554765 ), \\
     &  p_5 = ( 91.0664644166627, 49.0057636944973, 76.1125676676337, -9.92033815503652 ), \\ 
     &  p_6 = ( 197.326337775966, -3.11006048502754, 183.877222508616, -71.5344542606618 ), \\
     &  p_7=  ( 166.917065951017, -124.860731594604, 109.723100607740, -15.2297670494417 ),
\end{split} 
\ee
\end{small}
and define a normalized primitive amplitude through the following equation 
\be
{\cal A}^{LL,RR}_{ijk}({\lambda_i},{\lambda_j},{\lambda_k}; 
{\lambda_e},{\lambda_\mu}) 
= \frac{i}{(4 \pi)^2}\; \times 10^{-9} \times \tilde {\cal A}_{ijk}^{LL,RR}({\lambda_i},{\lambda_j},{\lambda_k}; 
{\lambda_e},{\lambda_\mu}).
\ee
The results for certain  helicity combinations of  
gluons and leptons are given in Table~\ref{table2}. We emphasize that diagrams where one $Z$-boson 
is emitted by decay products of another $Z$-boson, see Fig.~\ref{feyndia}(b), are {\it not} 
included in our calculation. The result for the amplitude squared and summed over 
colors and helicities of gluons and leptons was checked against the results of the OpenLoops 
program \cite{openloops} for a large number of kinematic points.\footnote{We are indebted to J.~Lindert
for making this comparison possible.} Finally, 
we note that the evaluation of the amplitude squared, summed over color and helicities, 
takes about 0.1 seconds per
phase-space point, making our implementation adequate for phenomenological needs.

In the context of NLO QCD computations, the process $gg \to ZZ +g $ represents an inelastic contribution. This 
 inelastic contribution  should be integrated  over all energies and angles of the 
emitted gluons,  including  the vanishingly small ones. 
Calculation of one-loop amplitudes for $gg \to ZZ g$ process  becomes unstable  if 
the gluon in the final state becomes soft or collinear to the collision axis. 
We deal with these instabilities 
by switching to quadruple precision where appropriate. 
To obtain the $gg \to ZZ$ cross section through 
NLO QCD,  we combine elastic and inelastic contributions using the 
$q_t$-subtraction  \cite{cagr} and, as a cross-check, the FKS subtraction~\cite{fks} methods.
The results that we present in the next Section are obtained by combining computations performed 
using the two subtraction schemes.

\begin{figure}[t]
\includegraphics[width=0.45\textwidth]{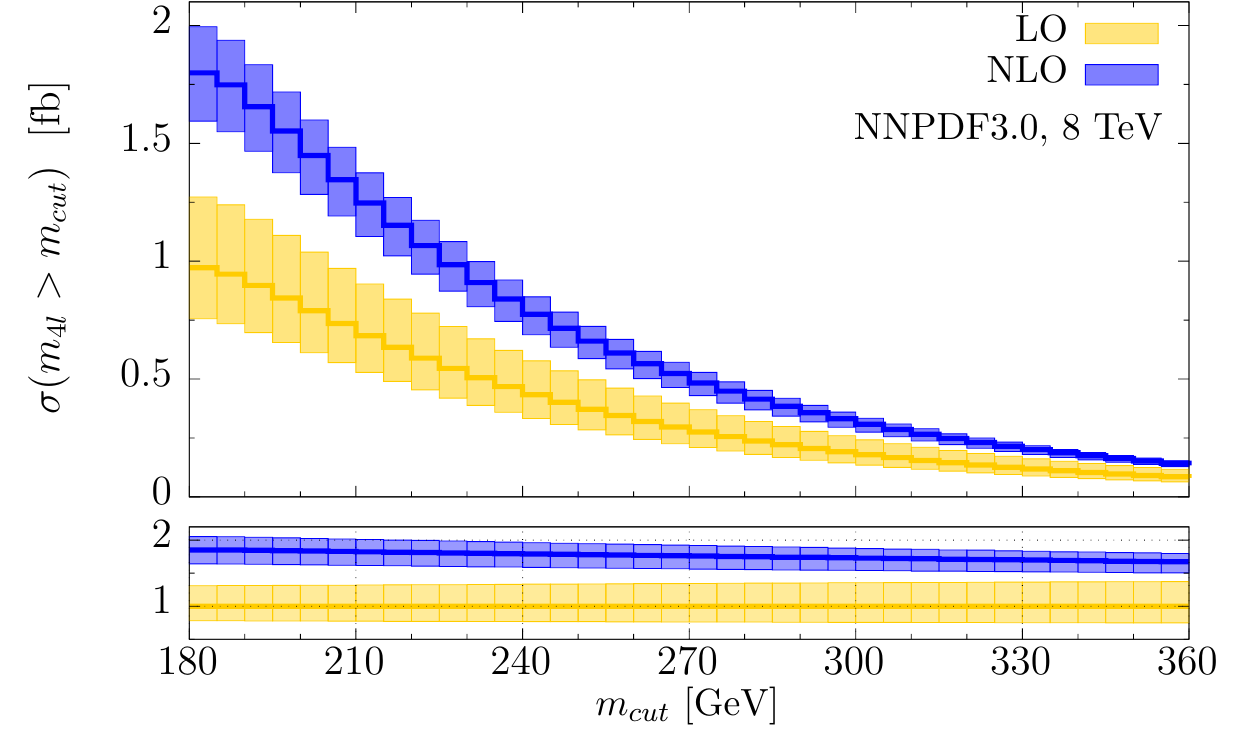}
\includegraphics[width=0.45\textwidth]{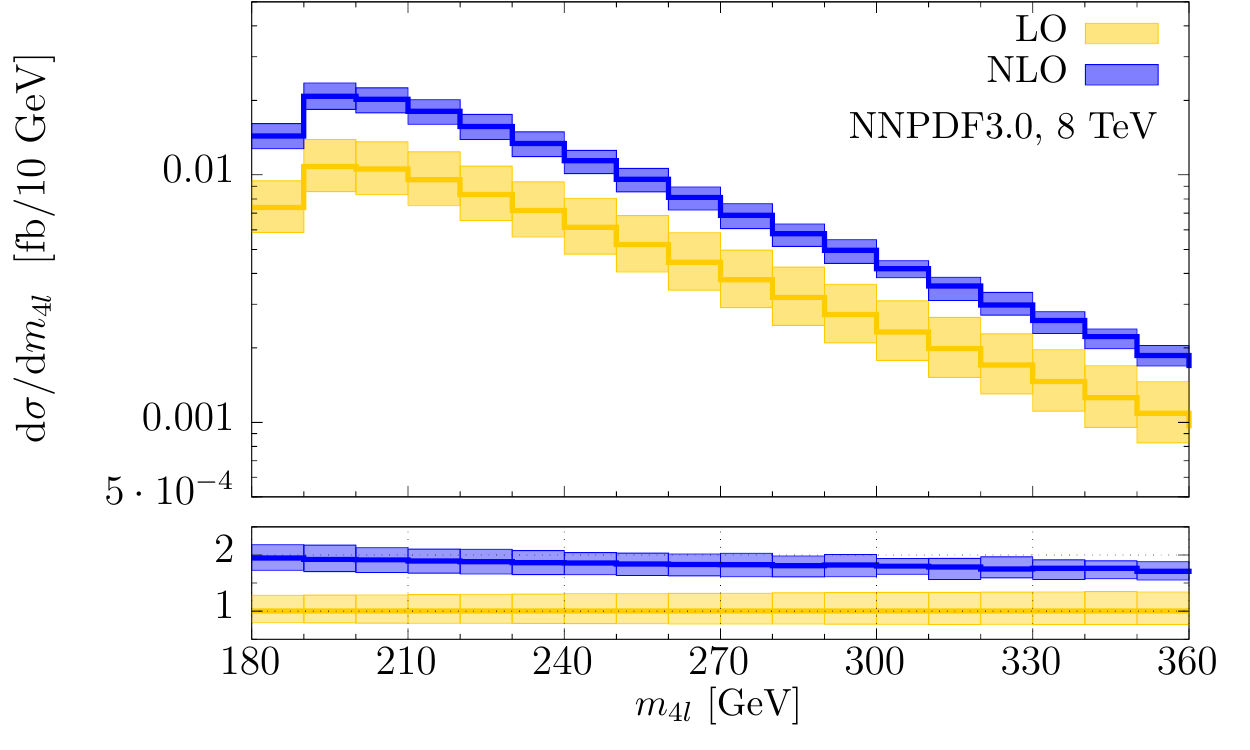}
\includegraphics[width=0.45\textwidth]{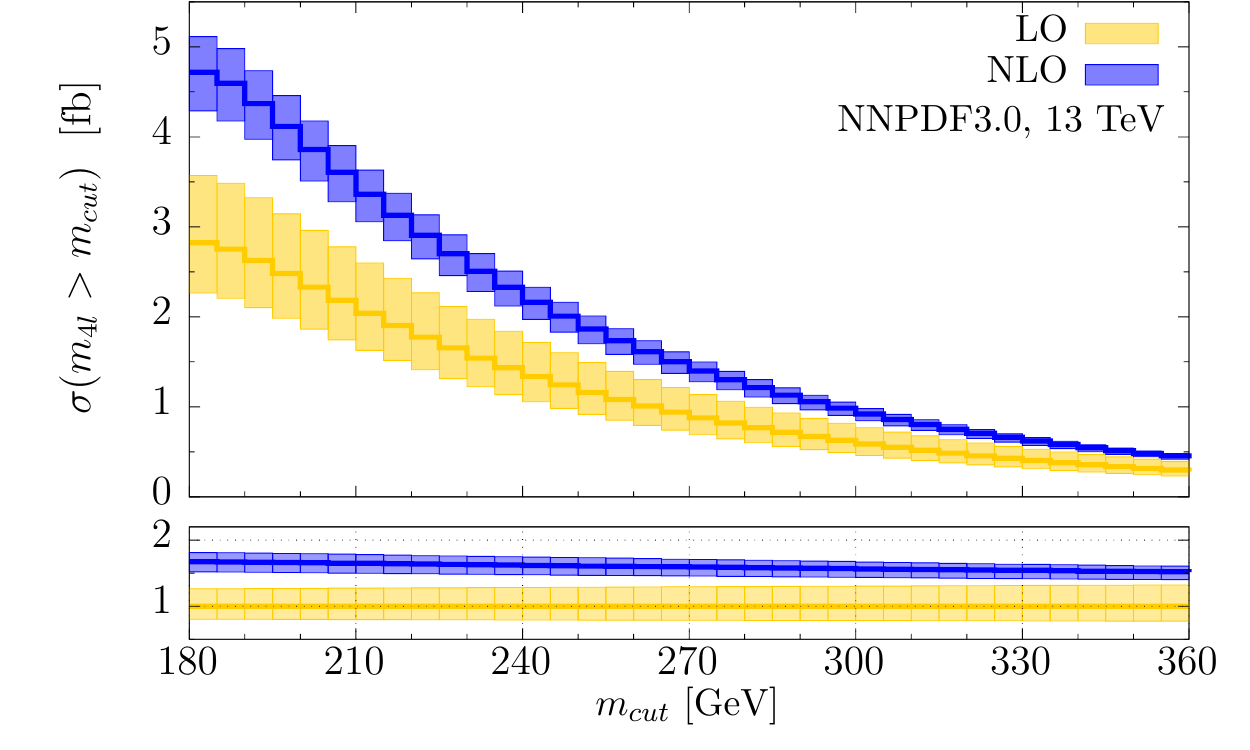}
\includegraphics[width=0.45\textwidth]{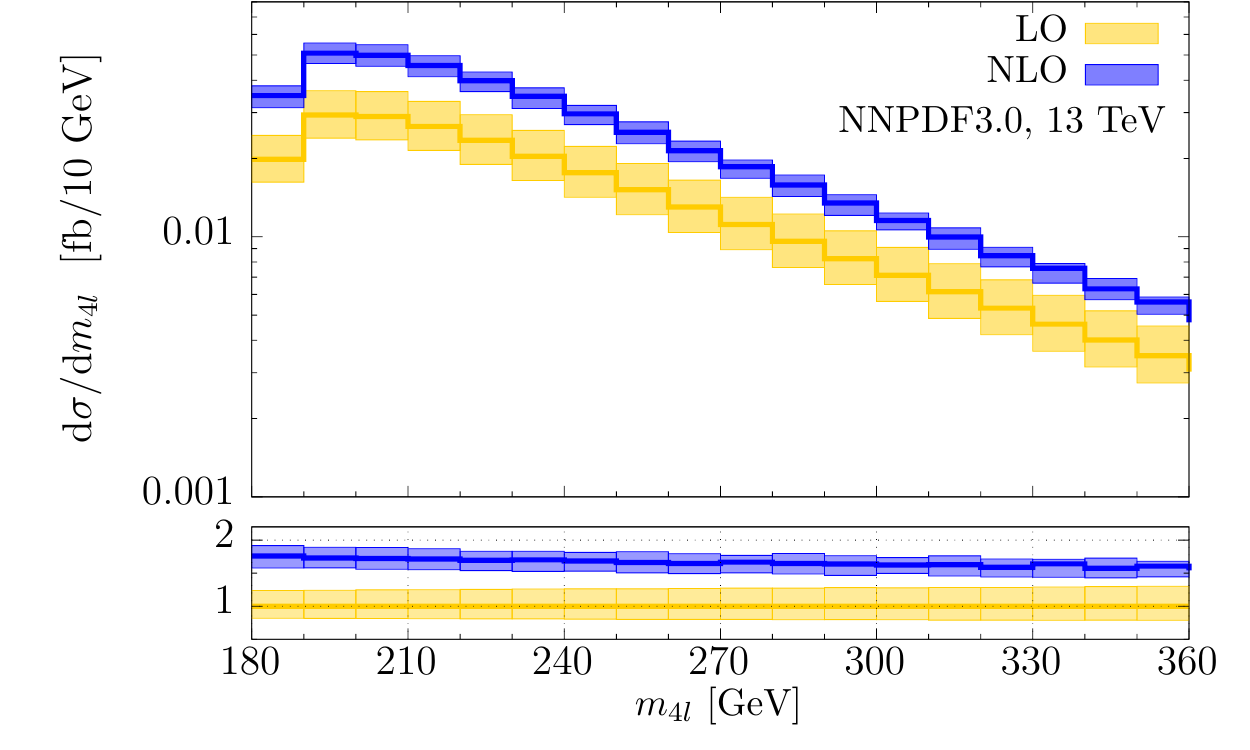}
\caption{Up, left: cumulative cross section for $gg \to (Z/\gamma)(Z/\gamma) \to e^+e^-\mu^+ \mu^-$ 
at the $8~{\rm TeV}$ LHC as a
function of the lower cut on four-lepton invariant mass. Up, right: 
distribution of the invariant mass of the four leptons in 
the reaction $ gg \to (Z/\gamma)(Z/\gamma) \to e^+e^- \mu^+ \mu^-$ at the $8~{\rm TeV}$ LHC. Lower 
panes show  ratios of the LO (yellow) and NLO (blue) distributions evaluated at three different scales to the LO 
distribution evaluated at $\mu = 2 m_Z$. Low: same as above for the $13~{\rm TeV}$ LHC.}\label{fig1}
\end{figure}

\section{Numerical results} 
\label{section3}
In this Section we  present the results of the calculation.  
We consider the process $gg \to (Z/\gamma)(Z/\gamma) \to e^+e^- \mu^+ \mu^-$ 
at the LHC.\footnote{We remind the reader that we only include double resonant diagrams, see Fig.\ref{feyndia}(a). Single resonant 
diagrams Fig.\ref{feyndia}(b) are only relevant for far off-shell production. They can be obtained by appropriate modifications of the $gg\to Zg$ 
amplitudes, see e.g.~\cite{Gehrmann:2013vga}.}
We generate invariant masses of $Z$ bosons around $m_Z$, using Breit-Wigner distributions. We require the $e^+e^-$ and $\mu^+\mu^-$ pair 
to have invariant masses $m_{l\bar l}\in (60,120)~{\rm GeV}$.
We use  leading (next-to-leading)
order parton distribution functions and the 
strong coupling constant for one- and two-loop calculations, respectively.  
We employ the NNPDF3.0 set of parton distribution functions and obtain the relevant 
values of the strong coupling constant from NNPDF routines \cite{Ball:2014uwa}. 

We begin with presenting the results for the total cross sections at the $8$~TeV LHC. We find 
\be
\sigma_{\rm LO}^{gg \to ZZ}  = 0.97^{+0.3}_{-0.2}~{\rm fb},
\;\;\;\;
\sigma_{\rm NLO}^{gg \to ZZ} = 1.8^{+0.2}_{-0.2}~{\rm fb},
\label{eq3.1}
\ee
where the central values refer  to the renormalization and 
factorization scales set to $\mu = 2 m_Z$ and the upper (lower) 
values to $\mu = m_Z$ ($\mu = 4 m_Z$).   
It follows from Eq.(\ref{eq3.1}) that QCD corrections to $gg \to ZZ$ are large -- 
the NLO cross section increases the LO cross section by ${\cal O}(60\% - 110\%)$, 
depending on the renormalization scale. 
For $\mu = 2 m_Z$, the cross section increases by  
$85\%$.

A similar situation occurs at the $13$~TeV LHC. We find 
\be
\sigma_{\rm LO}^{gg \to ZZ}  = 2.8^{+0.7}_{-0.6}~{\rm fb},
\;\;\;\;
\sigma_{\rm NLO}^{gg \to ZZ} = 4.7^{+0.4}_{-0.4}~{\rm fb}.
\label{eq3.13}
\ee
The NLO QCD corrections to $gg \to ZZ$ at $13~{\rm TeV}$ LHC are again 
significant but somewhat smaller than corrections at $8~{\rm TeV}$. Indeed, for the central value 
of the renormalization and factorization scales $\mu = 2 m_Z$, the  cross section increases by  $67\%$. 
For other values of the renormalization and factorization scales, the cross section increases by 
${\cal O}(40\%-90\%)$.

The large size of the QCD corrections is  reminiscent 
of the large QCD corrections to Higgs production in gluon fusion $gg \to H$ \cite{shiggs}. 
In addition, similar to the Higgs production case, the  scale variation of the leading order cross section 
provides a poor estimate  of the magnitude of next-to-leading order corrections \cite{shiggs}. 
We note that if  we take proximity of radiative effects in $gg \to ZZ$ and $gg \to H$ seriously, we  
should probably take 
$\mu = (2 m_Z)/2 = m_Z$ as the 
scale for which  higher-order radiative 
corrections to $gg \to ZZ$ will most likely be small. Thus, our best estimates of 
$gg \to ZZ$ contributions 
to $pp \to ZZ$  production cross section at $8$~TeV and 13~TeV LHC  are 
\be
\sigma_{pp \to ZZ}^{gg}(8~{\rm TeV}) = 2.0(2){\rm fb},\;\;\;\;
\sigma_{pp \to ZZ}^{gg}(13~{\rm TeV}) = 5.1(4){\rm fb}.
\label{eq.best}
\ee

Our results have important  implications 
for  the recently computed NNLO QCD corrections to $pp \to ZZ$ \cite{Cascioli:2014yka,Grazzini:2015nwa}
 at the $8$~TeV LHC. 
In that  case, the NNLO QCD corrections computed at  the scale $\mu = m_Z$ turned out to be 
close to $15\%$.  However,  a significant fraction --  $60\%$  of the total  NNLO QCD correction --
is due to the  leading order contribution  $gg \to ZZ$.  Our current computation shows that 
$gg \to ZZ$ receives large radiative corrections and the natural question is how these findings 
affect the central value of $pp \to ZZ$ cross section obtained  in 
Refs.~\cite{Cascioli:2014yka,Grazzini:2015nwa} and the theory uncertainty assigned to it. 

To answer this question,  we note that 
in Refs.~\cite{Cascioli:2014yka,Grazzini:2015nwa} the central scale was chosen 
to be $\mu = m_Z$ and that NNLO parton distribution functions 
  were used  for the calculation of $gg \to ZZ$ cross section.  Relative to our choices, 
the lower renormalization and factorization scale  increases the cross section while the choice of NNLO 
parton distribution functions 
makes the cross section smaller. We re-computed the LO $gg \to ZZ$ cross section using 
the setup of Ref.~\cite{Cascioli:2014yka} and compared it with our best value given 
in Eq.(\ref{eq.best}). We find that, to match our 
best prediction, the $8$~{\rm TeV} $gg \to ZZ$ cross section 
of  Ref.~\cite{Cascioli:2014yka} should be increased by about $80\%$.
In turn, this will lead to an increase in the total NNLO QCD correction to  $pp \to ZZ$ at $8$~TeV
from the current $12\%$, as calculated  in Ref.~\cite{Cascioli:2014yka}, to $18\%$.
This increase is beyond the ${\cal O}(3\%)$ scale variation of the NNLO QCD result for $pp \to ZZ$ 
used in Ref.~\cite{Cascioli:2014yka} to estimate the 
current uncertainty  in the theoretical prediction for  $pp \to ZZ$ cross section. 
Similar arguments also apply at the 13 TeV LHC. In this case the $16\%$
corrections quoted in Ref.~\cite{Cascioli:2014yka} would increase to approximately $23\%$. 

Next, we  consider kinematic distributions. We begin with  the invariant mass distribution of the 
four leptons produced in $gg \to ZZ$ shown in  Fig.~\ref{fig1}.  While radiative corrections 
are significant for all values of  $m_{4l}$,  they become smaller at higher values of four-lepton 
invariant masses. This is clearly seen  in Fig.~\ref{fig1} for both differential 
and cumulative\footnote{For different cuts on $m_{4l}$.} cross sections and for both the 8~TeV and
the 13~TeV LHC. 
This result is important for studies of the Higgs off-shell production where good understanding 
of the  shape of four-lepton invariant mass distribution is an important pre-requisite for 
constraining the Higgs width. Note that  for  $m_{4l}>2 m_t$  top-quark contributions,
neglected in our computation, become relevant.

\begin{figure}[t]
\includegraphics[width=0.45\textwidth]{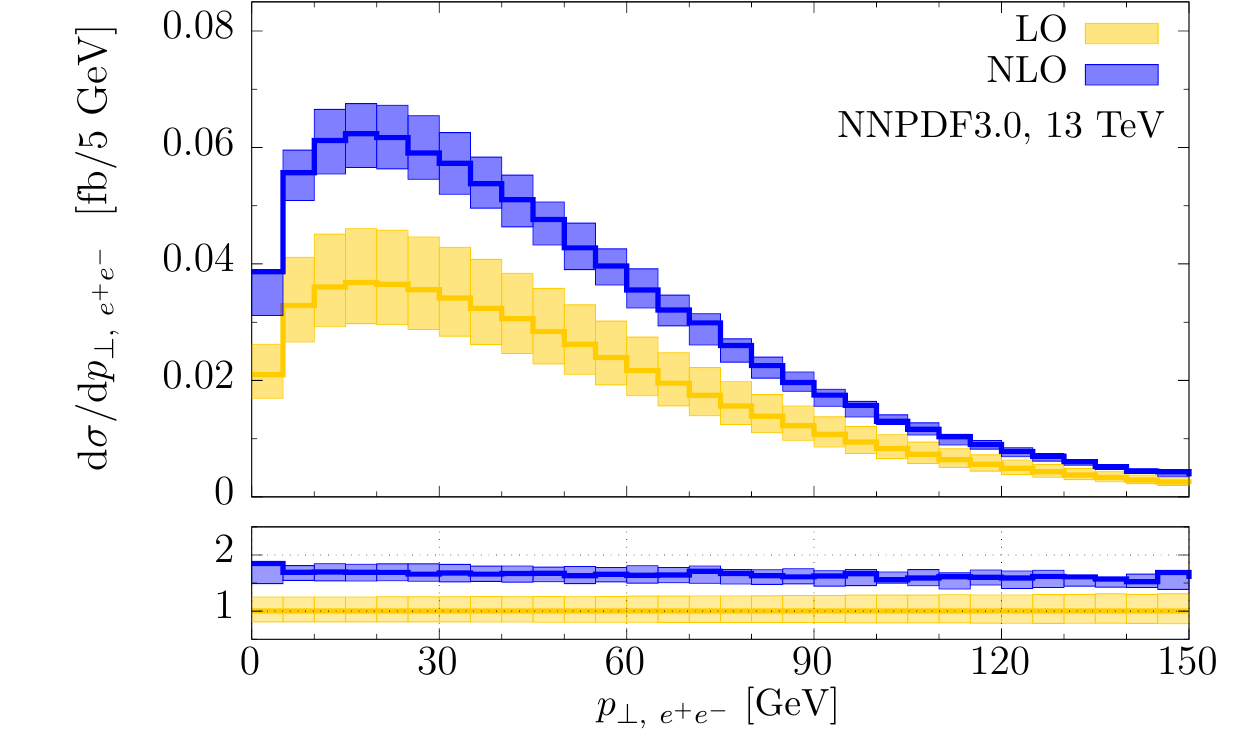}
\includegraphics[width=0.45\textwidth]{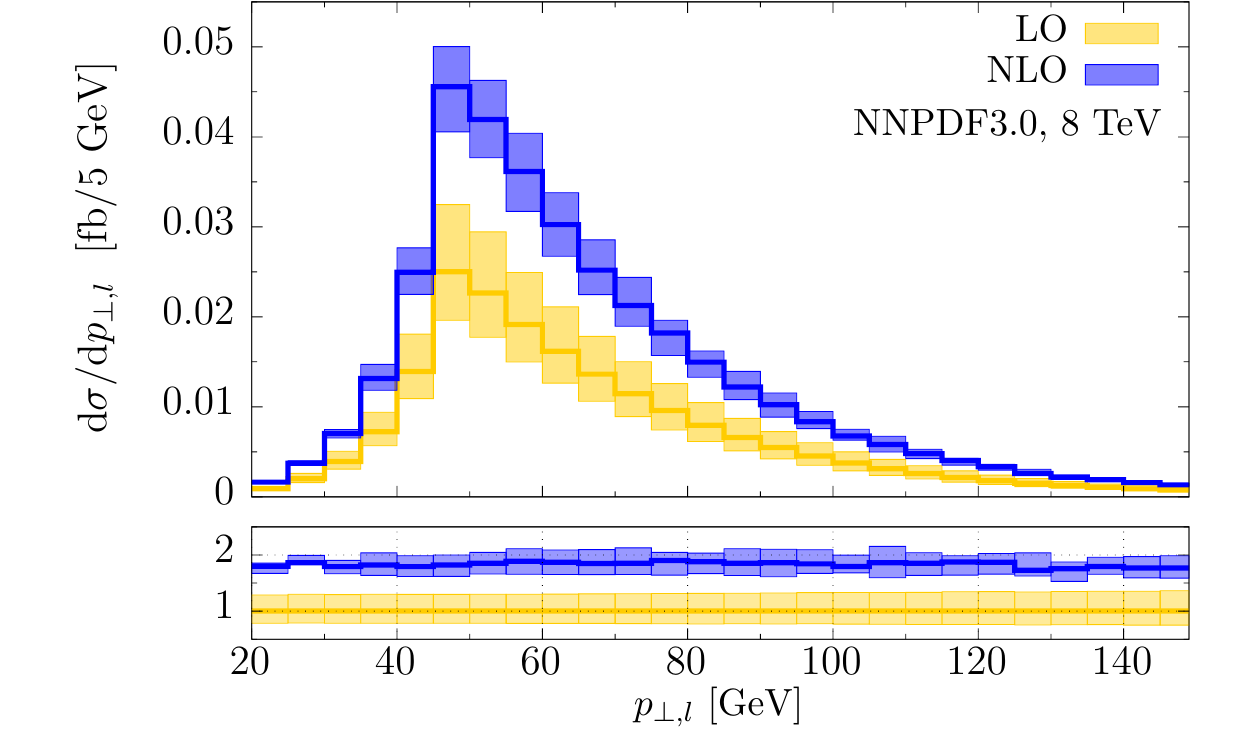}
\caption{Left:  transverse momentum distribution of an $e^+e^-$ pair  at the $13~{\rm TeV}$ LHC. 
Right: the hardest  lepton transverse momentum distribution at the $8~{\rm TeV}$ LHC. Lower 
panes show ratios of the LO (yellow) and NLO (blue) distributions evaluated at three different scales to the LO 
distribution evaluated at $\mu = 2 m_Z$. }\label{fig2}
\end{figure}

In Fig.~\ref{fig2} we show the transverse momentum distributions of the $e^+e^-$ pair and of the 
hardest lepton in the event. The QCD corrections to the transverse momentum distribution of 
the $e^+e^-$ pair decrease for large values of $p_{\perp,e^+e^-}$, similar to what is seen 
in the four-lepton invariant mass distribution. 
On the other hand, the QCD corrections for the transverse momentum distribution 
of the hardest lepton are independent  of the lepton $p_\perp$.

\section{Conclusions}
\label{section4}

In this paper we computed  QCD corrections to the  production of a pair of $Z$-bosons in 
gluon fusion through loops of massless quarks.  We  found  
that QCD corrections are large; they change the production cross section by almost a factor of two. 
These large QCD corrections are in line with expectations that transition of two 
gluons to a colorless final state  is strongly affected by QCD radiative effects;  QCD corrections 
of similar magnitude were observed earlier  in theoretical calculations 
of  $gg \to H$ \cite{shiggs} and $gg \to \gamma \gamma$ \cite{Bern:2002jx}
cross sections. 

Large QCD corrections to $gg \to ZZ$ are important for a number of reasons.  First, 
since the $gg \to ZZ$ process provides a significant fraction of the NNLO QCD contribution to 
$pp \to ZZ$,  our result 
suggests that existing theoretical predictions for $pp \to ZZ$ should be increased  by 
six to eight percent, depending on collider energy.  Since such an  
increase in the central value is outside the existing estimates of the residual 
theory uncertainty of  $pp \to ZZ$ cross section,  it will have  important 
consequences for ongoing  comparisons 
of experimental and theoretical results for  $pp \to ZZ$ at the LHC.
Second, 
good understanding  of $gg \to ZZ$ at high  four-lepton invariant masses is  crucial for  the so-called 
off-shell studies of the Higgs boson and, in particular, for the indirect determination of its width.  
The NLO QCD  calculation of $gg \to ZZ$ process 
allows us  to predict the $gg \to ZZ$ contribution to 
$pp \to ZZ$ cross section and kinematic distribution with 
the precision of about $10\%$; this    implies a residual theoretical 
uncertainty on $pp \to ZZ$ cross section  of just about two percent. Such a small uncertainty 
in the four-lepton production   cross section  is an essential  prerequisite for the success of forthcoming 
 off-shell studies of the Higgs boson, see a related discussion in Ref.~\cite{atlaswidth}. 

As a final comment, we note that our  calculation opens up a number of future research directions. Indeed, 
it is interesting to extend our calculation by combining massless and 
massive loop contributions to  $gg \to ZZ$ and  by including single resonant contributions and the interference 
of prompt $gg \to ZZ$ and $gg \to H^* \to ZZ$ amplitudes. This will allow us to  explore the region of four-lepton invariant 
masses both below the threshold of $ZZ$ production and at very high invariant masses.  We plan to do this in the near future. 

{\bf Acknowledgments} 
We are grateful to S. Pozzorini and, especially, to J. Lindert 
for their help in checking the  scattering amplitude for 
$gg \to ZZ+g$ process computed in this paper against the implementation in OpenLoops. 
We would like to thank R.K.~Ellis and J.~Campbell for useful conversations about computation 
of the rational part. 
F.C. and K.M. thank the Mainz Institute for Theoretical Physics (MITP) for
hospitality and partial support during the program \emph{Higher Orders and Jets for LHC}. 
This research is partially supported by BMBF grant 05H15VKCCA.
Fermilab is operated by Fermi Research Alliance, LLC
under Contract No. De-AC02-07CH11359 with the United States Department of Energy.


\begin{thebibliography}{99}

\bibitem{atlas7} G. Aad {\it et al.} ATLAS collaboration, Phys. Rev. D{\bf 87}, 112001 (2013) [Erratum 
ibid., {\bf 88}, 079906 (2013)].

\bibitem{cms7} 
  S.~Chatrchyan {\it et al.} [CMS Collaboration],
  Eur.\ Phys.\ J.\ C {\bf 73}, no. 10,  2610 (2013).

\bibitem{cms8} CMS Collaboration, CMS-PAS-SMP-12-013. 

\bibitem{Khachatryan:2014jba} 
  V.~Khachatryan {\it et al.} [CMS Collaboration],
  Eur.\ Phys.\ J.\ C {\bf 75}, no. 5,  212 (2015).

\bibitem{Khachatryan:2014kca} 
  V.~Khachatryan {\it et al.} [CMS Collaboration],
  Phys.\ Rev.\ D {\bf 92}, no. 1,  (2015), 012004.

\bibitem{Khachatryan:2014iha} 
  V.~Khachatryan {\it et al.}  [CMS Collaboration],
  Phys.\ Lett.\ B {\bf 736}, 64 (2014).

\bibitem{atlaswidth} 
  G.~Aad {\it et al.} [ATLAS Collaboration],
  Eur.\ Phys.\ J.\ C {\bf 75}, no. 7, 335  (2015).


\bibitem{Catani:2011qz} 
  S.~Catani, L.~Cieri, D.~de Florian, G.~Ferrera and M.~Grazzini,
  Phys.\ Rev.\ Lett.\  {\bf 108}, 072001  (2012).

\bibitem{Grazzini:2013bna} 
  M.~Grazzini, S.~Kallweit, D.~Rathlev and A.~Torre,
  Phys.\ Lett.\ B {\bf 731}, 204  (2014).

\bibitem{Cascioli:2014yka} 
  F.~Cascioli {\it et al.},
  Phys.\ Lett.\ B {\bf 735}, 311 (2014).

\bibitem{Gehrmann:2014fva} 
  T.~Gehrmann, M.~Grazzini, S.~Kallweit, P.~Maierhoefer, A.~von Manteuffel, S.~Pozzorini, D.~Rathlev and L.~Tancredi,
  Phys.\ Rev.\ Lett.\  {\bf 113}, no. 21, 212001 (2014).

\bibitem{Grazzini:2015nwa} 
  M.~Grazzini, S.~Kallweit and D.~Rathlev,
  JHEP {\bf 1507},  085  (2015).

\bibitem{Grazzini:2015hta} 
  M.~Grazzini, S.~Kallweit and D.~Rathlev,
  arXiv:1507.06257 [hep-ph].


\bibitem{Glover:1988rg} 
  E.~W.~N.~Glover and J.~J.~van der Bij,
  Nucl.\ Phys.\ B {\bf 321}, 561  (1989).

\bibitem{Glover:1988fe} 
  E.~W.~N.~Glover and J.~J.~van der Bij,
  Phys.\ Lett.\ B {\bf 219}, 488  (1989).

\bibitem{Dicus:1987dj} 
  D.~A.~Dicus, C.~Kao and W.~W.~Repko,
  Phys.\ Rev.\ D {\bf 36},  1570 (1987).


\bibitem{Binoth:2006mf}
  T.~Binoth, M.~Ciccolini, N.~Kauer and M.~Kramer,
  JHEP {\bf 0612},  046  (2006).

\bibitem{Bonvini:2013jha} 
  M.~Bonvini, F.~Caola, S.~Forte, K.~Melnikov and G.~Ridolfi,
  Phys.\ Rev.\ D {\bf 88}, no. 3,  034032 (2013).

\bibitem{Caola:2013yja} 
  F.~Caola and K.~Melnikov,
  Phys.\ Rev.\ D {\bf 88},  054024  (2013).

\bibitem{Campbell:2013una} 
  J.~M.~Campbell, R.~K.~Ellis and C.~Williams,
  JHEP {\bf 1404},  060 (2014).

\bibitem{Kauer:2012hd} 
  N.~Kauer and G.~Passarino,
  JHEP {\bf 1208},  116 (2012).

\bibitem{ellis} 
  J.~M.~Campbell, R.~K.~Ellis and C.~Williams,
  JHEP {\bf 1110}, 005  (2011).

\bibitem{Azatov:2014jga} 
  A.~Azatov, C.~Grojean, A.~Paul and E.~Salvioni,
  Zh.\ Eksp.\ Teor.\ Fiz.\  {\bf 147},  410 (2015);
  [J.\ Exp.\ Theor.\ Phys.\  {\bf 120}, 354 (2015)].


\bibitem{mcfm} J.~Campbell and R.K.~Ellis, http://mcfm.fnal.gov.

\bibitem{gg2VV} 
  N.~Kauer,
  JHEP {\bf 1312},  082 (2013).

\bibitem{Melnikov:2015laa} 
  K.~Melnikov and M.~Dowling,
  Phys.\ Lett.\ B {\bf 744}, 43 (2015).

\bibitem{Gehrmann:2013cxs}
  T.~Gehrmann, L.~Tancredi and E.~Weihs,
  JHEP {\bf 1308}, 070  (2013).
  
\bibitem{Gehrmann:2014bfa} 
T.~Gehrmann, A.~von Manteuffel, L.~Tancredi and E.~Weihs,
  JHEP {\bf 1406}, 032 (2014).

\bibitem{planar} 
  J.~M.~Henn, K.~Melnikov and V.~A.~Smirnov,
  JHEP {\bf 1405}, 090 (2014).

\bibitem{nonplanar} 
  F.~Caola, J.~M.~Henn, K.~Melnikov and V.~A.~Smirnov,
  JHEP {\bf 1409}, 043 (2014).

\bibitem{Papadopoulos:2014hla} 
  C.~G.~Papadopoulos, D.~Tommasini and C.~Wever,
  JHEP {\bf 1501}, 072 (2015).

\bibitem{Caola:2014iua}
  F.~Caola, J.~M.~Henn, K.~Melnikov, A.~V.~Smirnov and V.~A.~Smirnov,
  JHEP {\bf 1411},  041 (2014).

\bibitem{Gehrmann:2015ora} 
  T.~Gehrmann, A.~von Manteuffel and L.~Tancredi,
  arXiv:1503.04812 [hep-ph].

\bibitem{Caola:2015ila} 
  F.~Caola, J.~M.~Henn, K.~Melnikov, A.~V.~Smirnov and V.~A.~Smirnov,
  JHEP {\bf 1506}, 129 (2015).

\bibitem{vonManteuffel:2015msa} 
  A.~von Manteuffel and L.~Tancredi,
  JHEP {\bf 1506}, 197 (2015),

\bibitem{Agrawal:2012df} 
  P.~Agrawal and A.~Shivaji,
  Phys.\ Rev.\ D {\bf 86}, 073013 (2012).

\bibitem{Campanario:2012bh} 
  F.~Campanario, Q.~Li, M.~Rauch and M.~Spira,
  JHEP {\bf 1306}, 069 (2013).


\bibitem{openloops} 
  F.~Cascioli, P.~Maierhofer and S.~Pozzorini,
  Phys.\ Rev.\ Lett.\  {\bf 108}, 111601 (2012).

\bibitem{madloop} 
  V.~Hirschi and O.~Mattelaer,
  arXiv:1507.00020 [hep-ph].


\bibitem{bern} Z. Bern, L.J.~Dixon and D.A.~Kosower, Nucl. Phys. {\bf B513}, 3 (1998).
\bibitem{britto} R.~Britto, F.~Cachazo and B.~Feng, Nucl. Phys. {\bf B725}, 275  (2005).
\bibitem{opp} G.~Ossola, G.G.~Papadopoulos, R.~Pittau, Nucl. Phys. {\bf B763}, 147 (2007).

\bibitem{Ellis:2007br} 
  R.~K.~Ellis, W.~T.~Giele and Z.~Kunszt,
  JHEP {\bf 0803}, 003 (2008).

\bibitem{me} W.T.~Giele, Z.~Kunszt and K. Melnikov, JHEP {\bf 04} , 049 (2008).

\bibitem{Ellis:2011cr} 
  R.~K.~Ellis, Z.~Kunszt, K.~Melnikov and G.~Zanderighi,
  Phys.\ Rept.\  {\bf 518}, 141 (2012).

\bibitem{Henn:2014yza} 
  J.~M.~Henn and J.~C.~Plefka,
  Lect.\ Notes Phys.\  {\bf 883}, 1 (2014).


\bibitem{cagr} S.~Catani and M.~Grazzini, Phys. Rev. Lett. {\bf 98}, 222002  (2007).

\bibitem{Catani:2013tia}
  S.~Catani, L.~Cieri, D.~de Florian, G.~Ferrera and M.~Grazzini,
  Nucl.\ Phys.\ B {\bf 881}, 414 (2014). 
  

\bibitem{direct} S.~Badger, JHEP {\bf 0901}, 049  (2009).

\bibitem{Campbell:2014gua} 
  J.~M.~Campbell, R.~K.~Ellis, E.~Furlan and R.~Rontsch,
  Phys.\ Rev.\ D {\bf 90}, no. 9, 093008 (2014).

\bibitem{Ellis:2008qc} 
  R.~K.~Ellis, W.~T.~Giele, Z.~Kunszt, K.~Melnikov and G.~Zanderighi,
  JHEP {\bf 0901}, 012 (2009).

\bibitem{Melia:2010bm} 
  T.~Melia, K.~Melnikov, R.~Rontsch and G.~Zanderighi,
  JHEP {\bf 1012}, 053 (2010).

\bibitem{Melia:2011dw} 
  T.~Melia, K.~Melnikov, R.~Rontsch and G.~Zanderighi,
  Phys.\ Rev.\ D {\bf 83}, 114043 (2011).

\bibitem{Melia:2012zg} 
  T.~Melia, K.~Melnikov, R.~Rontsch, M.~Schulze and G.~Zanderighi,
  JHEP {\bf 1208}, 115 (2012).

\bibitem{badger} S.D.~Badger, E.W.N.~Glover and V.~Khoze, JHEP {\bf 0601}, 066 (2006).

\bibitem{fks}
  S.~Frixione, Z.~Kunszt and A.~Signer,
  Nucl.\ Phys.\ B {\bf 467}, 399 (1996).

\bibitem{Gehrmann:2013vga} 
  T.~Gehrmann, L.~Tancredi and E.~Weihs,
  JHEP {\bf 1304}, 101 (2013).

\bibitem{Ball:2014uwa} 
 R.~D.~Ball {\it et al.} [NNPDF Collaboration],
  JHEP {\bf 1504}, 040 (2015).

\bibitem{shiggs} Graudenz, M. Spira and P. M. Zerwas, Phys. Rev. Lett. {\bf 70}
, 1372 (1993);
S. Dawson, Nucl. Phys. B {\bf 359}, 283 (1991);
A. Djouadi, M. Spira and P. M. Zerwas, Phys. Lett. B {\bf 264}, 440 (1991);
M. Spira, A. Djouadi, D. Graudenz and P. M. Zerwas, 
Nucl. Phys. B {\bf 453}, 17 (1995);
R. V. Harlander and W. B. Kilgore, Phys. Rev. Lett. {\bf 88}, 201801 (2002);
C. Anastasiou and K. Melnikov, Nucl. Phys. B {\bf 646}, 220 (2002);
V. Ravindran, J. Smith and W. L. van Neerven, Nucl. Phys. B {\bf 665}, 325 (2003);
C.~Anastasiou, C.~Duhr, F.~Dulat, F.~Herzog and B.~Mistlberger,
 Phys.\ Rev.\ Lett.\  {\bf 114}, 212001 (2015).

\bibitem{Bern:2002jx} 
  Z.~Bern, L.~J.~Dixon and C.~Schmidt,
  Phys.\ Rev.\ D {\bf 66}, 074018 (2002).











\end{thebibliography}
\end{document}